# Scaling of graphene field-effect transistors supported on hexagonal boron nitride: radio-frequency stability as a limiting factor

Pedro C. Feijoo[*,1,2], Francisco Pasadas[1], José M. Iglesias[3], María J. Martín[3], Raúl Rengel[3], Changfeng Li[2], Wonjae Kim[2,4], Juha Riikonen[2], Harri Lipsanen[2], David Jiménez[1]

[1] Departament d'Enginyeria Electrònica, Escola d'Enginyeria, Universitat Autònoma de Barcelona, 08193 Bellaterra, Spain

[2] Department of Electronics and Nanoengineering, Aalto University, 02150 Espoo, Finland

[3] Departamento de Física Aplicada, Universidad de Salamanca, Salamanca 37008, Spain

[4] VTT Technical Research Center of Finland, 02150 Espoo, Finland

* *Corresponding author:* <u>PedroCarlos.Feijoo@uab.cat</u>

**Abstract**

The quality of graphene in nanodevices has increased hugely thanks to the use of hexagonal boron nitride as a supporting layer. This paper studies to which extent hBN together with channel length scaling can be exploited in graphene field-effect transistors (GFETs) to get a competitive radio-frequency (RF) performance. Carrier mobility and saturation velocity were obtained from an ensemble Monte Carlo simulator that accounted for the relevant scattering mechanisms (intrinsic phonons, scattering with impurities and defects, etc.). This information is fed into a self-consistent simulator, which solves the drift-diffusion equation coupled with the two-dimensional Poisson's equation to take full account of short channel effects. Simulated GFET characteristics were benchmarked against experimental data from our fabricated devices. Our simulations show that scalability is supposed to bring to RF performance an improvement that is, however, highly limited by instability. Despite the possibility of a lower performance, a careful choice of the bias point can avoid instability. Nevertheless, maximum oscillation frequencies are still achievable in the THz region for channel lengths of a few hundreds of nanometers.

Supplementary data available

## 1. Introduction

Graphene promises to stand out as a channel material in analog radio-frequency (RF) electronics due to its two-dimensional (2D) character and carrier transport properties: a mobility up to $2 \cdot 10^5$ cm$^2$ V$^{-1}$ s$^{-1}$ and a high saturation velocity of $4 \cdot 10^7$ cm s$^{-1}$ [1–5]. Besides, graphene presents a remarkably high mechanical strength, with an elastic modulus of up to 550 N m$^{-1}$ and a breaking strength of up to 35 N m$^{-1}$ [6,7], which also makes it a feasible channel material for flexible electronics. Nevertheless, the transport properties decrease significantly when graphene is the active part of a substrate supported device because of the charge scattering [1,8]. Hexagonal boron nitride (hBN), however, with an atomically perfect surface relatively free of dangling bonds and charge traps, has proved to be an exceptional dielectric for graphene field-effect transistors (GFETs) [9]. Graphene devices fabricated on hBN exhibit up to one order-of-magnitude improvement in mobility and carrier inhomogeneities in comparison with conventional oxide dielectrics [10].

In the effort to successfully develop the next-generation of GFET supported on hBN technology, device simulation tools must describe accurately both the electrostatics and carrier transport across the graphene taking into account the specificity of the graphene/hBN interaction. An appropriate description of the carrier transport requires the inclusion of the relevant scattering mechanisms that drive the carrier mobility and saturation velocity. In a previous work [11] we developed a self-consistent model that solved the drift-diffusion transport equation coupled with the 2D Poisson's equation in order to study the short-channel effects (SCE) in GFETs, that is, to analyze the influence of the drain voltage on the electrostatic field and charge channel distribution in short channel GFETs. However, both mobility and saturation velocity were considered constant, so the model ignored the complexity of the interplay among the scattering processes that, in fact, strongly depend on carrier density. In this paper, we have considered the various scattering mechanisms, including the specific influence of the hBN substrate on the carrier transport. For this purpose, carrier transport properties were calculated by a self-consistent ensemble Monte Carlo (EMC) simulator for graphene that allows to assess the role played by each type of scattering [12,13]. This information is fed into the self-consistent GFET model to obtain current-voltage (I-V) characteristics. Additionally, we have formulated a small-signal model of the GFET composed by parameters extracted from linearization around a bias point of direct current (DC) simulations. Different from what the studies ever reported so far, the proposed model guarantees charge conservation and assumes non-reciprocal capacitances, which is essential for accurate calculations of the transistor RF figures of merit (FoM). These kind of models are required to bridge the gap between both device and circuit levels and to make comparisons with other existing RF technologies, e.g. those based on Si or III-V compounds [14]. Importantly, our GFET simulator and corresponding small-signal model can deal with SCE, which significantly reduce the expected cutoff ($f_T$) and maximum oscillation ($f_{max}$) frequencies [11] with respect to the simple projection derived from the long-channel behavior. Also, we have investigated the stability of a GFET when it operates as an amplifier using the small-signal model combined with microwave-engineering techniques. We specifically look into the dependence of stability on both channel length and graphene quality. Lack of stability can prevent a transistor from working as an amplifier within the targeted RF window, so it is desirable to know the conditions that ensure stability and to estimate the possible trade-off with the power gain.

First, this article briefly describes the drift-diffusion based simulator. Then, we discuss the methodology used to extract the mobility and saturation velocity taking full account of the intrinsic scattering mechanisms, the presence of impurities and defects across the graphene sheet, and the interactions with the substrate and top gate dielectric. Next, the small-signal model is presented, emphasizing the relevance of guaranteeing charge conservation. The simulator has been benchmarked against DC experimental results from our fabricated devices. Using this model we have investigated the scalability of GFETs supported on hBN targeting RF applications. This has been carried out with assumption of different scenarios for both impurities and defects concentrations. The final part of this work thoroughly studies the RF stability of GFETs.

## 2. Methods

### 2.1 Device structure and self-consistent simulator

We studied the dual-gate GFET represented in figure 1, which is based on a structure where a graphene layer is encapsulated between thin layers of hBN and $Al_2O_3$. The hBN, with a thickness of the order of tens of nanometers, acts as the supporting intermediate layer on the substrate while the $Al_2O_3$ layer plays the role of the top gate insulator. The $SiO_2$/n-type Si substrate is also utilized as a back gate stack. We have neglected the influence of the thin hBN layer in the electrostatics calculations since its contribution to the back gate capacitance is relatively small. In fact, the error in the gate capacitance calculation is lower than 3% for a 10 nm thick hBN layer with a relative permittivity of 5 [15]. However, we do have considered its influence on the electronic transport through the calculation of low-field mobility $\mu_{LF}$ and velocity saturation $v_{sat}$. Regarding the electronic transport, the drift-diffusion mechanism was considered as an appropriate description, although this assumption can only be applied if the mean free path (MFP) of carriers remains much shorter than the channel length. We will see that it is the case for the considered scaled devices in this study under room temperature operation.

GFETs were simulated using the method described ref. [11], which solves Poisson's equation and drift-diffusion transport equation in a self-consistent way. The dashed rectangle in figure 1 remarks the active area of the device and corresponds to the domain where the Poisson's equation is resolved. The simulator uses both the $\mu_{LF}$ and $v_{sat}$ extracted by the Monte Carlo methodology discussed in section 2.2. Also, it obtains the stationary values of the drain current $I_{ds}$ and the carrier distribution along the channel as a function of voltages applied to the gate, back gate and drain terminals with respect to the source ($V_{gs}$, $V_{bs}$ and $V_{ds}$, respectively). More details of the simulator can be found in section S1 of the supplementary data.

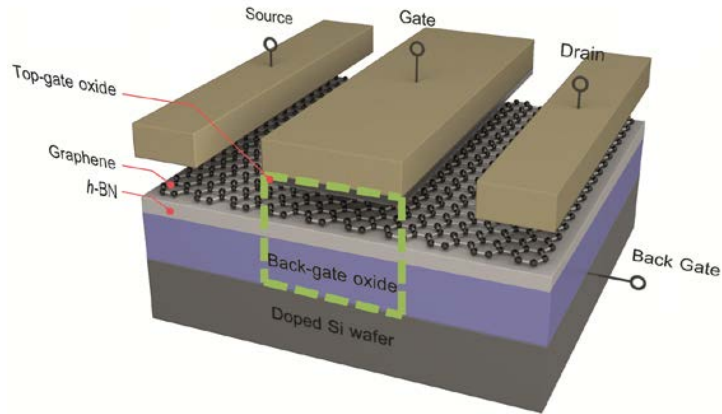

**Figure 1** Schematic of the GFET on hBN/SiO$_2$/Si substrate that is used in simulations and experiments.

*2.2 Scattering mechanisms and carrier transport properties from Monte Carlo simulations*

The values of $\mu_{LF}$ and $v_{sat}$ as a function of the carrier density have been obtained for different levels of impurities and defects in graphene by means of an EMC simulator [12,13]. The model includes optical and acoustic intravalley phonons, electron-electron interactions, impurity scattering and scattering with defects, together with scattering with remote surface polar phonons (SPP) from the substrates. In the case of acoustic and optical phonons the scattering rates are obtained according to the deformation potential approximation [16], fitting the parameters to reproduce the rates provided by the first-principles density functional theory [17]. Short range carrier-carrier interactions are implemented by considering a static Coulomb screened potential model [18]. Impurity scattering is derived from the 2D Fourier transform of the electrostatic potential with charged centers [19]. Regarding graphene defects, both point defects and dislocations are considered as a single scattering mechanisms thanks to the approximation described in ref. 27 with a defect parameter $\alpha$ accounting for the average defect potential, the effective potential range and the density of defects. Both $\alpha$ and the density of impurities $n_{imp}$ were taken as fitting parameters to reproduce the experimental mobility. The model allows for the Pauli Exclusion Principle to treat the effect of degeneracy. From the point of view of these EMC simulations, the graphene layer is placed between the top and the bottom dielectrics. The influence of the top gate metal (separated from the graphene layer by a relatively thin dielectric of thickness $t_t$) on the screening of impurities, remote phonons and carrier-carrier scattering has been incorporated by modifying the dielectric function with a suitable Green function [21,22], although its effect for the gate oxide considered is negligible. On the other hand, the bottom substrate is considered thick enough so that the effects on screening on the back gate can be neglected. An in-depth discussion of the EMC model is presented in the supplementary data.

The number of simulated carriers ranged from $10^5$ to $10^6$ depending on the carrier concentration, and the time step considered in the simulation to update the distribution functions for the self-consistent treatment of the Pauli Exclusion Principle equaled 2.5 fs. The saturation velocity was determined at 20 kV cm$^{-1}$, as done by other authors in experimental works [23]. From the velocity-field curves obtained from the EMC simulation it is possible to extract $\mu_{LF}$ and $v_{sat}$ as a function of the carrier concentration $n_s$. Different scenarios defined by the parameters $\alpha$ and $n_{imp}$ have been considered. Specifically, we have assumed three levels of graphene quality referred in this work as the high, the intermediate mobility scenario and the low mobility scenarios corresponding to 1, 2 and 3 in figures 2(a) and (b), respectively. In order to reproduce the experimental extracted mobility (close to 1500 cm$^2$ V$^{-1}$ s$^{-1}$) of our experimental devices, we used a $n_{imp}$ of $8 \cdot 10^{12}$ cm$^{-2}$ and $\alpha$ equal to 0.10 eV nm, and this case represents the scenario 3. The extracted mobility is also similar to that reported in ref. 27. The combination of $\alpha$ and $n_{imp}$ of the scenario 2 represents the fitting to experimental mobility data for graphene on SiO$_2$ [23], and is included for comparison purposes (an improved quality sample in a realistic case). We also include the ideal scenario 1 where neither impurities nor defects are present, which defines the best possible graphene quality.

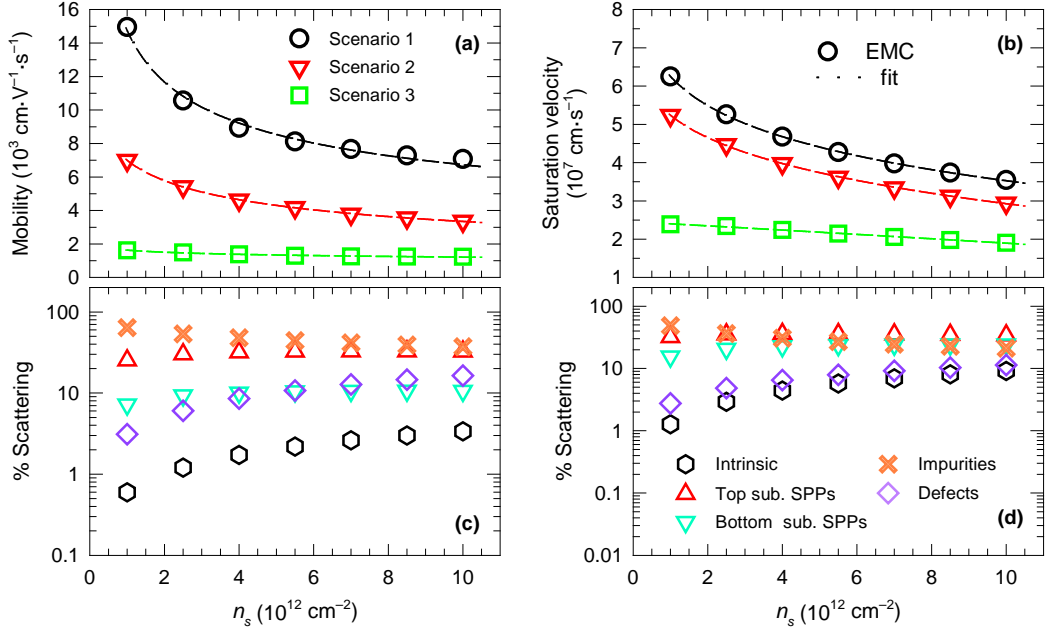

**Figure 2** Carrier mobility (a) and saturation velocity (b) in graphene supported on hBN as a function of the carrier concentration for different combinations of impurity densities and defects. Scenario 1 corresponds to an ideal situation, without neither impurities nor defects; scenario 2 to $n_{imp} = 0.95 \cdot 10^{12}$ cm$^{-2}$ and $\alpha = 0.07$ eV nm; and scenario 3 to $n_{imp} = 8 \cdot 10^{12}$ cm$^{-2}$ and $\alpha = 0.1$ eV nm. Symbols represent the EMC data while lines represent the fit to the analytical expression. For scenario 2 (intermediate mobility), the percentage of each scattering type as a function of the carrier concentration is represented in the case of low field (c) and high field (d).

The EMC results allow determining which scattering mechanism becomes more critical to establish the values of $\mu_{LF}$ and $v_{sat}$. As an example, the percentage of scattering mechanisms is shown in figures 2(c) and (d) for the intermediate mobility scenario; the results for the extreme mobility cases are also shown in figure S2 of the supplementary data. In the low mobility scenario 3, the dominant scattering type are impurities, with a secondary role of defects at high carrier concentrations. Scattering with SPP from the top oxide is also relevant, particularly at large $n_s$, having more importance in setting the saturation velocity value. In the scenario 2, impurities are still dominant at low fields, but interactions with the top and bottom dielectrics play an increasingly important role, especially at high electric fields, for which they can even become the primary scattering type. Finally, in the ideal scenario 1, the low field mobility and saturation velocity are mainly influenced by interactions with SPP of the top (primarily) and bottom dielectrics, with progressive larger influence of intrinsic phonons at high carrier concentrations.

Once that the EMC data for $\mu_{LF}$ and $v_{sat}$ in each scenario are obtained, the information must be introduced in the self-consistent GFET simulator. Both $\mu_{LF}$ and $v_{sat}$ vs. $n_s$ have been fitted to different mathematical expressions, as shown in Table S1 of ESM. In this way, the effect of the carrier scattering activity on the RF performance can be readily evaluated.

As a final note, it is relevant to discuss the validity of the drift-diffusion transport used in our work, so a calculation of the MFP is necessary. EMC simulator allows obtaining their values, which are presented in figure S3 of the supplementary data together with the ratio between the gate length $L$ and the MFP in each case, as a function of the $L$. This ratio is the relevant feature in order to discuss the diffusive character of transport and the validity of the drift-diffusion model. The results are presented for two carrier concentrations, $10^{12}$ and $10^{13}$ cm$^{-2}$. As it can be observed, the MFP ranges from a few nanometers to more than 200 nm depending on the carrier concentration and the electric field; however, the validity of the diffusive model is guaranteed since MFP is always significantly smaller than the featured gate length for each case.

### 2.3 Charge-based small-signal model of GFET and derived RF performance

In order to assess the RF performance of the GFET, the device can be conveniently considered as a two-port network in the common source configuration, as depicted in figure 3(a). At the input port, a small-signal alternate current (AC) voltage source $v_S$, of internal admittance $Y_S$, transfers power and current to the network, being both $v_S$ and $Y_S$ complex numbers (phasors). A load admittance $Y_L$ is connected at the output to get the transferred power. A small-signal model in form of an extrinsic admittance matrix $Y$ describes the behavior of the two-port network and its analysis provides both the device RF performance and stability information. The small-signal parameters $y_{ij}$ of $Y$ can be extracted from the linearization of current and charge DC characteristics (see section S2 in the supplementary data). The $Y$-parameters depend strongly on (i) the frequency ($\omega = 2\pi f$), (ii) the bias point: $V_{gs}$ and $V_{ds}$, and (iii) the series resistances of the gate, source and drain terminals ($R_G$, $R_S$ and $R_D$, respectively). A full justification of the high frequency model can be found in ref. 30.

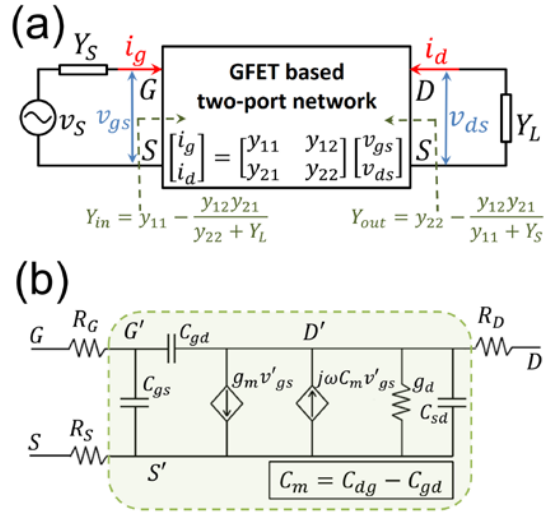

**Figure 3** (a) GFET conceptualized as a two-port network, characterized by its $Y$ matrix, connected to source and load admittances. (b) Small-signal model of a three-terminal GFET for high frequency analysis.

A charge control model is needed to determine the small-signal parameters. We apply a Ward-Dutton's linear charge partition scheme, which is charge conserving,

The intrinsic small-signal model of the GFET, within the shaded region of figure 3(b), is suitable for high frequency analysis [24,25]. A distinctive feature of this model is that no *a* *priori* assumption on the reciprocity of intrinsic capacitances has been made. This kind of assumptions are usually made for silicon transistors and have often been imported directly to GFET intrinsic capacitance modeling [26–29]. In our opinion, this practice is not well justified and important deviations of the calculated RF FoMs has been reported in ref. 30.

The RF FoMs can be extracted from $Y(\omega)$. The extrinsic cutoff frequency $f_{T,x}$ is defined from the current gain, defined as $\beta(\omega) = - i_d / i_g$, that is, the ratio of the output current $i_d$ and the input current $i_g$. The magnitude of $\beta(\omega)$ presents a maximum value when the output is shorted ($|Y_L| \rightarrow \infty$) and its expression takes the following form [31]:

$$\beta_{max}(\omega) = \frac{-y_{21}}{y_{11}} \tag{1}$$

By definition, $f_{T,x}$ is the frequency at which the current gain is equal to one, that can be written as:

$$\left| \beta_{max}\left(2\pi f_{T,x}\right) \right| = 1 \tag{2}$$

Before discussing the procedure to get $f_{max}$, we must recall the concept of stability of a general two-port amplifier circuit in terms of the $Y$-parameters of a transistor. Stability guarantees that no adventitious oscillations can appear at a network for any passive source and load admittances connected to the input and output ports, respectively. This requires that the reflection coefficients of the input and output ports are smaller than one. Equivalently, the stability of a network can be assessed by means of the $K$-$\Delta$ test, which is based on the evaluation of the two following factors [31]:

$$K(\omega) = \frac{2 \, \mathrm{Re}(y_{11}) \, \mathrm{Re}(y_{22}) - \mathrm{Re}(y_{12} y_{21})}{|y_{12} y_{21}|} \tag{3a}$$

$$\Delta(\omega) = \frac{(Y_0 - y_{11})(Y_0 - y_{22}) - y_{12}y_{21}}{(Y_0 + y_{11})(Y_0 + y_{22}) - y_{12}y_{21}} \tag{4b}$$

where $Y_0 = 1/ Z_0$ is the characteristic admittance and $Z_0$ is the characteristic impedance (usually taken as 50 Ω). Both conditions $K > 1$ and $|\Delta| < 1$ are necessary and sufficient to ensure device stability. In this case, any passive load and input admittance provide a stable behavior of the network. Selecting an optimum set of $Y_S$ and $Y_L$, an optimized power gain can be obtained, referred as the maximum available gain (MAG). However, if $-1 < K < 1$, the network is said to be conditionally stable, that is, it becomes stable only for certain combinations of $Y_S$ and $Y_L$. Among those combinations that provide stability, the maximum attainable power gain is known as the maximum stable gain (MSG). Therefore, the maximum gain $|G(\omega)|_{max}$ can be calculated from $Y$ and it depends on the value of the stability factor $K$ following these relations:

$$|G(\omega)|_{max} = \begin{cases} MAG = \left|\frac{y_{21}}{y_{12}}\right| \left(K - \sqrt{K^2 - 1}\right) & K \geq 1; \; |\Delta| < 1 \\ MSG = \left|\frac{y_{21}}{y_{12}}\right| & -1 < K < 1 \end{cases} \tag{5}$$

Once $G(\omega)$ has been defined, $f_{max}$ can be calculated as follows:

$$|G(2\pi f_{max})|_{max} = 1 \tag{6}$$

The small-signal model considered ignores the electron inertia, that is, the non-quasi-static effects, which are expected to be relevant when GFETs operate near or beyond $f_T$. Then, the quasi-static approach only provides a rough estimate of the RF FoMs. A more refined calculation should include the effect of kinetic inductance [32]. Ignoring the kinetic inductance means that the quasi-static model should not be used for operating frequencies beyond $f_T$. Nevertheless, using a low frequency model to derive both $f_T$ and $f_{max}$ is consistent with the usual practice of extrapolating $f_T$ and $f_{max}$ from the measurement of current and power gains at low frequencies. Anyway, the main objective of this paper consists in analyzing the relation between stability and scalability, which hasn't been discussed yet in the context of 2D material technologies. Other aspect not included in the model is hot electron effects. Ref. [33] showed that impact ionization produced by hot carriers produces an "up-kick" in the I-V characteristics, prompting a worsening of $g_d$. This means that the FoMs predicted by our model should be regarded as an upper limit.

### 2.4 Fabrication and measurement of GFETs

The hBN flakes were prepared by mechanical exfoliation on 285 nm SiO$_2$/Si substrate. To fabricate a GFET, graphene grown by photo-thermal chemical vapor deposition [34] was transferred onto the hBN and, subsequently, was patterned with oxygen plasma to define the graphene channel. Ti/Au (2/50 nm) metal electrodes were contacted to the graphene channel utilizing lift-off technique. Afterwards, 26-nm-thick Al$_2$O$_3$ was deposited by atomic layer deposition on the top surface of the structure for a top gate dielectric insulator. A second layer graphene was transferred on the dielectric layer and patterned for a gate electrode. Finally, Ti/Au metal lead for a gate was fabricated. All DC measurements were performed at room temperature using semiconductor parameter analyzer (HP4155A) in ambient conditions.

## 3. Results and discussions

### 3.1 Comparison with experimental values

We have checked the predictive behavior of the self-consistent simulator by comparing its outcome with experimental I-V curves of a GFET. The graphene channel is on the top of a layer of hBN with a thickness of the order of tens of nm, which is placed on 285 nm thick

SiO$_2$. An atomic force microscopy (AFM) image of the device and its structure layout including the hBN flake are shown in figures 4(a) and (b), respectively. For simulation purposes, we have considered the parameters of the GFET presented in Table 1. The source and drain contact resistances have been assumed as one fitting parameter, $R_S = R_D$, to optimize the agreement between the model outcome and the experiment. Note that the contact resistance is sensitive to the back gate voltage since, under the source/drain contacts, graphene is overlapped with the global back gate [35,36]. The relation between both DC intrinsic voltages ($V_{gs}$ and $V_{ds}$) and extrinsic voltages ($V_{gs,ext}$ and $V_{ds,ext}$ for gate-to-source and drain-to-source, respectively) depend on drain current $I_{ds}$, $R_S$ and $R_D$ as follows:

$$V_{ds,ext} = V_{ds} + I_{ds}(R_S + R_D) \tag{7a}$$

$$V_{gs,ext} = V_{gs} + I_{ds}R_S \tag{7b}$$

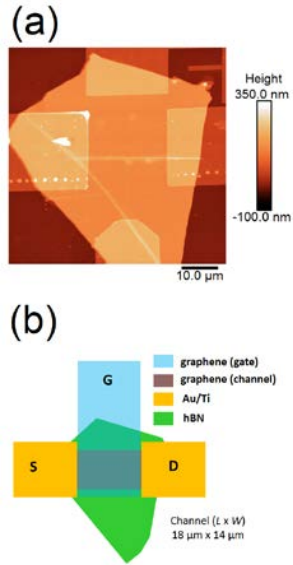

**Figure 4** (a) AFM image and (b) device layout of the experimental GFET supported on hBN.

**Table 1** Parameters of the nominal GFET supported on hBN used in the simulation.

| Parameter | Value |
|---|---|
| Channel length $L$ | 18 µm |
| Channel width $W$ | 14 µm |
| Top insulator thickness $t_t$ | 26 nm |
| Back insulator thickness $t_b$ | 285 nm |
| Top insulator relative permittivity $\varepsilon_t$ | 9 |
| Back insulator relative permittivity $\varepsilon_b$ | 3.9 |
| Top gate flatband voltage $V_{gs0}$ | -2.5 V |
| Back gate flatband voltage $V_{bs0}$ | 0 V |
| $n_{imp}$ | $8 \cdot 10^{12}$ cm$^{-2}$ |
| $\alpha$ | 0.10 eV nm |

Figure 5 plots together the experimental and calculated curves, showing good agreement even if very different back gate voltages $V_{bs}$ are considered. The values of the series resistances found for each $V_{bs}$ resulted from contact resistance modulated by the back gate. A thorough discussion of this effect can be found in refs. 41 and 40. We neglected the influence of interface traps that might exist at the insulator-graphene interface. Nevertheless, in figure S3 of the supporting data, we report on the impact that might exist due to the non-ideal $Al_2O_3$/graphene interface [37] with an interface trap capacitance of 10 fF µm$^{-2}$. We have found that this additional capacitance, which is a realistic value, does not significantly affect the I-V characteristics nor the transcapacitances.

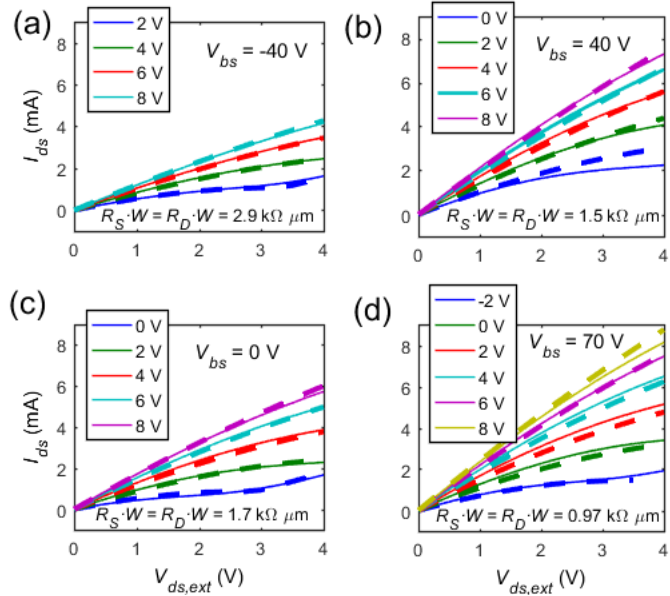

**Figure 5.** Comparison of simulated (dashed lines) and experimental (solid lines) I-V characteristics of the nominal device for different back gate voltages: (a) $V_{bs}$ = -40 V, (b) 0 V, (c) 40 V and (d) 70 V. The legends indicate the $V_{gs,ext}$.

### 3.2 RF performance scaling and device stability

The small-signal model has been used to investigate the scalability of RF performance via channel length reduction, considering device stability at the same time. We anticipate that stability plays a vital role, especially in short-channel transistors. We have studied the effect of graphene quality for the two extreme mobility scenarios (1 and 3) considered in the Monte Carlo simulations. The scenario 2 gives medium RF performance and its discussion would be qualitatively similar, so we have not included it for the sake of brevity. To run the simulations we have used state-of-the art values of the source/drain series resistance $R_S \cdot W = R_D \cdot W = 200 \ \Omega \ \mu m$ [38]. The gate series resistance was calculated considering a metal gate contacted on both sides of the device [24]. If the GFET width is large enough, we can approximate the gate resistance to be inversely proportional to channel length. For instance, a realistic value of $R_G \cdot L$ could be 4.4 $\Omega$ $\mu m$, calculated assuming a 60 nm thick wolfram gate (resistivity of 56 nΩ m). The increase of $R_G$ with scaling compromises the ultimate $f_{max}$ of GFETs so gate resistance minimization is key in RF applications [39]. Regarding the density of puddles, we have assumed it as zero for the results presented in this paper. However, we have not found significant deviations in the RF performances assuming puddle densities up to ~$10^{11}$ cm$^{-2}$, provided that the chosen bias point is far enough from the Dirac point. A more in-depth investigation of the puddle concentration and its effects on RF performances can be found in figure S5 of the supplementary data.

The GFET RF performance is, in general, dependent on the bias point. For our RF investigation, we have chosen the combination $V_{ds} = 0.6$ V and $V_{gs}$ - $V_D = 2$ V, where $V_D$ is the Dirac voltage, so the device is biased in the region where $f_{max}$ and $f_{T,x}$ are quite insensitive to $V_{gs}$ (see figure S6 of supplementary data). This eases the comparison with other devices since the performance does not depend so much on the bias point.

The impact of the channel length downscaling and graphene quality on the power gain can be observed in figure 6(a). The MSG presents a slope of 10 dB dec$^{-1}$ for all devices, and reducing the channel length by a factor of 2 results in an increase of the power gain of 5.4 dB while $f_{max}$ grows from 6.5 to 16 GHz. A similar improvement is achieved if the electron mobility increases from the scenario 3 up to scenario 1.

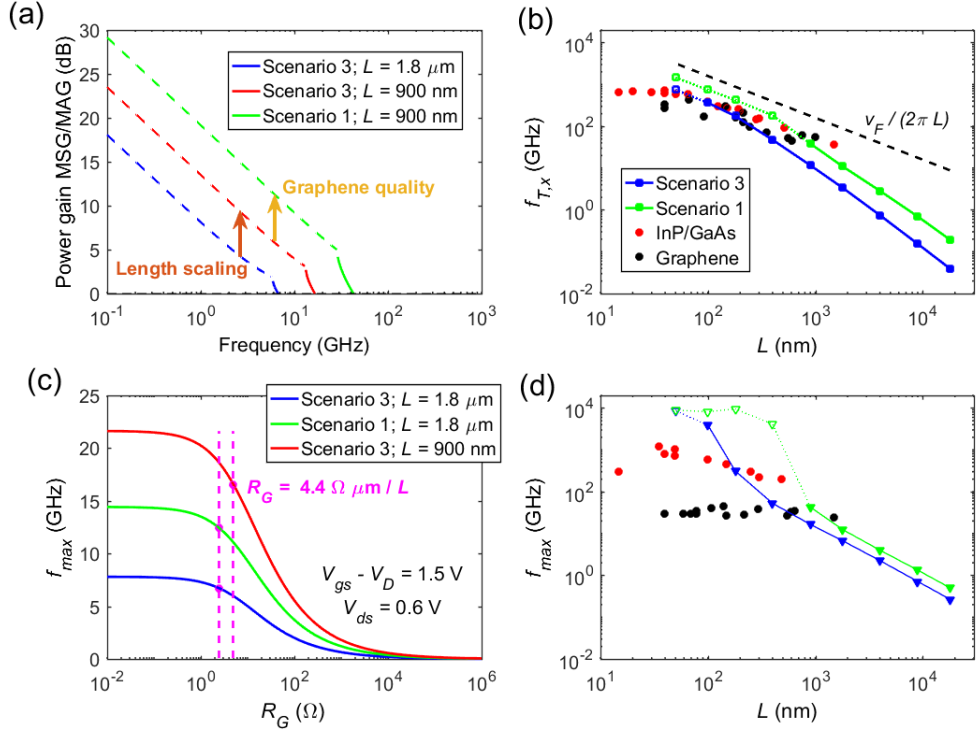

**Figure 6** (a) Calculated MSG (dashed lines) and MAG (solid lines) of GFETs. We have considered channel lengths of 1.8 µm and 900 nm with and without defects and impurities. Scaling of $f_{T,x}$ (b) and $f_{max}$ (d) comparing the simulations of GFET on hBN (green and blue symbols) with experimental results from state-of-the art GFET on conventional dielectrics (black symbols) and InP or GaAs transistors (red symbols). For the GFET on hBN different graphene quality scenarios are considered. Closed and open symbols mean stable and unstable devices, respectively. Instability implies that the GFET amplifier is unusable at this particular bias point. In (b), the dashed line corresponds to the physical limit of the $f_{T,x}$, that is, $v_F / (2\pi L)$, which comes out from the minimum possible transient time in a graphene channel $L / v_F$. The chosen bias point was $V_{gs}$ - $V_D$ = 2 V and $V_{ds}$ = 0.6 V. (c) $f_{max}$ as a function of the gate resistance $R_G$, considering different channel lengths and graphene qualities. The circles refer to the point where $R_G \cdot L$ is equal to 4.4 Ω µm.

Next, we have shown in figures 6(b) and (d) the scaling of $f_{T,x}$ and $f_{max}$ with $L$. Details on the scaling of the small-signal parameters can be found in figure S7 of supplementary data. For long channel lengths (> 1 µm), $f_{T,x}$ scales as $1/L^2$. This is because $g_m$ is proportional to $1/L$ while $C_g$ scales as $L$. However, for short channels (< 1 µm), the scaling law approaches $1/L$ because of saturation velocity effects, which make $g_m$ quite insensitive to $L$. Increasing the graphene quality improves $f_{T,x}$ in a factor larger than 2. The numbers shown in figure 6(b) are comparable to what has been reported value for InP and GaAs high electron mobility transistors (HEMTs), which are the highest reported value for RF transistors [2]. Importantly, we have found out that for short-channel devices, the device becomes unstable at very short $L$, especially in the high mobility scenario. This particular issue of instability will be later discussed.

Regarding $f_{max}$ scalability, shown in figure 6(d), we have found a different trend to the one for $f_{T,x}$. At long channel lengths, the scaling law is $1/L^n$ with $1 < n < 2$, which is in fact a scaling power smaller than $f_{T,x}$ due to the upscaling of $R_G$. The increase in graphene quality slightly improves $f_{max}$. However, at short channel lengths (< 1 μm), there is a huge increase in $f_{max}$ because of current saturation driven by the velocity saturation effect. Our simulations indicate the great potential that the GFET on hBN might have to push $f_{max}$ into the THz region. Intrinsic I-V curves for different devices depending on the quality of graphene and channel length can be found in figure S8 of Supplementary Data.

The gate series resistance is in fact an important source of RF performance degradation. figure 6(c) illustrates how $f_{max}$ decreases with the gate series resistance. The graph compares the results for devices with different channel lengths (900 nm and 1.8 μm) and devices with different levels of graphene quality (mobility scenarios 1 and 3). It is then clear that minimizing the gate series resistance produces an important improvement in $f_{max}$, even more prominent when the channel becomes shorter. Besides, no relation has been found between the stability of the GFET and the gate resistance.

Let us turn now the attention to the device stability issue. In figure 7 we have plotted the stability factors $K$ and $|\Delta|$ considering different channel lengths and graphene qualities.

First, we focus on the stability of devices with different $L$, shown in figures 7(a) and (b). While longer devices show conditional stability, the factor $K$ in the short channel case ($L = 180$ nm) decreases below -1 for a set of frequencies between ~$10^2$ and $10^4$ GHz. The scaled transistor thus enters in the unstable region, which prevents it from working properly as a power amplifier. Similarly, for a fixed gate length, figures 7(c) and (d) show that moving towards the ideal mobility scenario could be detrimental as the device is more prone to instability. This could imply sacrificing some power gain to restore stable RF operation. Therefore, we suggest that a low mobility scenario is helpful to extend the device stability to shorter channel lengths, although paying the prize of getting smaller RF performance. Device stability is eventually lost when the transistor with the low mobility is aggressively scaled down to 40 nm.

Finally, it is relevant to analyze how the choice of the drain bias could impact on the RF performance and device stability. Figure 7(e) compares $f_{T,x}$ and $f_{max}$ scaling at both $V_{ds} = 0.6$ V and 0.4 V. For long channel lengths ($L > 1$ μm), the reduction in $V_{ds}$ gives a slight decrease in the FoMs. However, a reduced $V_{ds}$ can be used to extend the device stability to lower channel lengths down to 180 nm. As a result, we can conclude that the choice of the bias point is important not only to maximize $f_{T,x}$, $f_{max}$ or the power gain but also to make sure that the device works in the stable region.

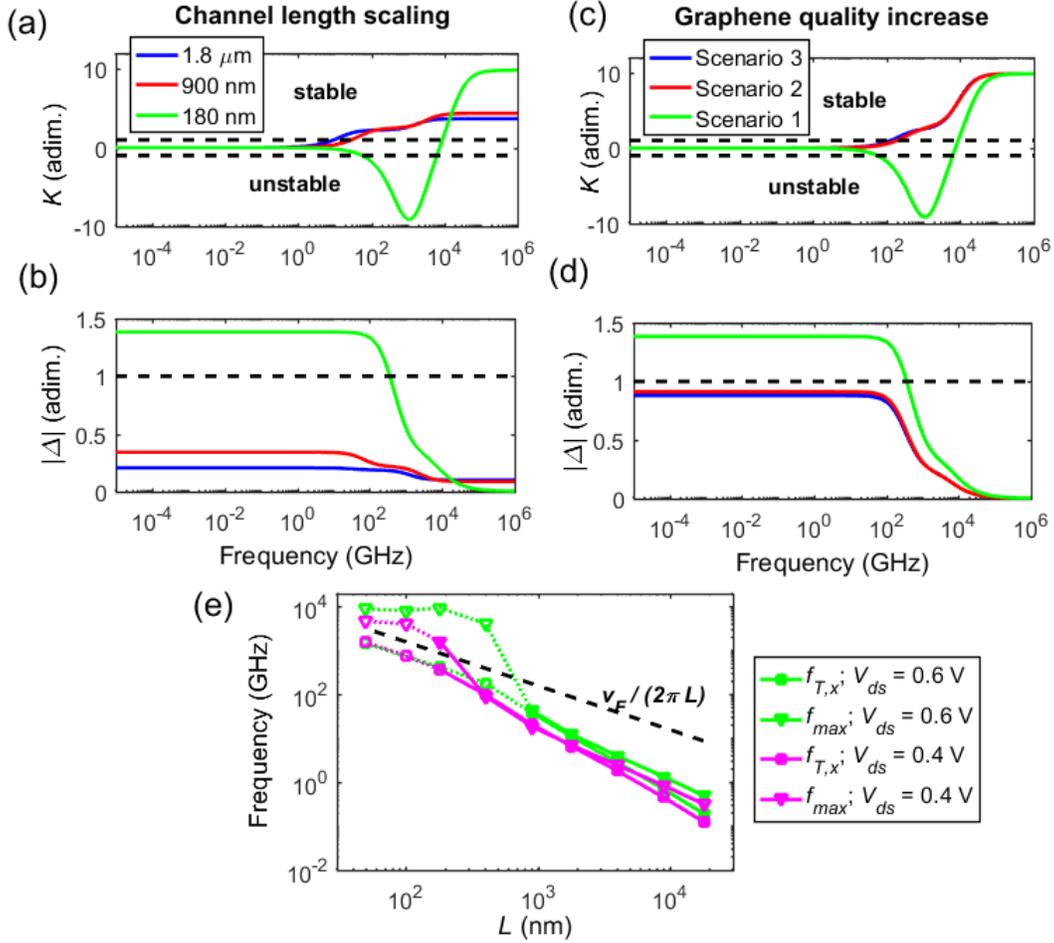

**Figure 7** (a) and (b) *K-Δ* stability test considering the effect of the channel length scaling and (c) and (d) the effect of graphene quality. Dashed lines separate the regions of instability and stability. In the *K*-frequency plot, the middle region ($|K| < 1$) corresponds to conditional stability. (e) Scaling of $f_{T,x}$ and $f_{max}$ at different drain voltages (with the same $V_{gs}$ - $V_D$ = 2 V). Closed and open symbols indicate stability and instability, respectively. The dashed line corresponds to the physical limit for the $f_{T,x}$.

## 4. Conclusions

We have analyzed the electrical behavior of the GFET supported on hBN. For such a purpose, we have simulated transistors following a multiscale approach, which combines (i) EMC simulations to study electron transport in graphene and (ii) self-consistent GFET simulations in a drift-diffusion scheme to include SCE. Monte Carlo simulations account for the most relevant scattering mechanisms that affect carrier transport in real samples and they provide the mobility and saturation velocity dependence with the carrier concentration. The results have been successfully compared with experimental measurements of fabricated devices. Furthermore, we have studied the RF performance with a charge-conserving small-signal model of the GFET. Graphene quality and channel length scaling could be ways to improve the RF performance up to THz and beyond. However, our results indicate that short-channel GFETs with very high mobility may be unstable, and therefore, not usable. The bias point is also important to guarantee a stable operation, so a careful design is needed of both the device and the bias voltages used. Finally, we have proved that increasing $f_{max}$ to competitive values requires a low gate series resistance, especially in the case of short-channel devices.

### Acknowledgements

We acknowledge the provision of facilities and technical support by Aalto University at Micronova Nanofabrication Centre for GFET fabrication. This work is funded by the European Union's Horizon 2020 research and innovation program (grant agreement No 696656), the *Ministerio de Economía y Competitividad* (projects TEC2013-42622-R, TEC2015-67462-C2-1-R, TEC2016-80839-P, MINECO/FEDER and grant FJCI-2014-19643), the *Ministerio de Educación* (CAS16/00043) and the *Generalitat de Catalunya* (project 2014 SGR 384).

Supplementary data

# Scaling of graphene field-effect transistors supported on hexagonal boron nitride: radio-frequency stability as a limiting factor

Pedro C. Feijoo[1,2], Francisco Pasadas[1], José M. Iglesias[3], María J. Martín[3], Raúl Rengel[3], Changfeng Li[2], Wonjae Kim[2,4], Juha Riikonen[2], Harri Lipsanen[2], David Jiménez[1]

[1] Departament d'Enginyeria Electrònica, Escola d'Enginyeria, Universitat Autònoma de Barcelona, 08193 Bellaterra, Spain
[2] Department of Electronics and Nanoengineering, Aalto University, 02150 Espoo, Finland
[3] Departamento de Física Aplicada, Universidad de Salamanca, Salamanca 37008, Spain
[4] VTT Technical Research Center of Finland, 02150 Espoo, Finland

## S1. Self-consistent simulator

We present here a brief account of the self-consistent simulator used to calculate the electrical behavior of dual-gate GFETs, as depicted in figure 1 of the main text. Details of the simulator can be found thoroughly described in ref. [1]. The bias voltages applied to the electrodes of top gate, drain and back gate with respect to the source ($V_{gs}$, $V_{ds}$ and $V_{bs}$, respectively), induce a sheet charge density $\sigma_{\text{free}}(y) = q[p(y) - n(y)]$ in the graphene sheet. Here, $p$ and $n$ are the hole and electron concentration in graphene, $q$ is the elementary charge and $y$ is the axis that goes from source ($y = 0$) to drain ($y = L$), where $L$ is the channel length. This charge distribution is needed in turn to calculate the electrostatic potential $\psi(x,y)$ inside the GFET by means of the Poisson's equation. Figure S1 shows the 2D domain where this equation is solved, where $x$ is the position along the axis that goes from back to top gate. Assuming that the GFET width $W$ (in the $z$ direction) is large as compared with the other dimensions of the device, the Poisson's equation can be written as follows:

$$\nabla \cdot [\varepsilon_r(x,y)\varepsilon_0 \nabla\psi(x,y)] = \rho_{\text{free}}(x,y) \tag{S1}$$

where $\varepsilon_0$ is the vacuum dielectric constant, and $\varepsilon_r(x,y)$ is the relative dielectric constant, which is equal to $\varepsilon_t$ and $\varepsilon_b$ inside the top and back dielectrics, respectively, and $\varepsilon_G$ in graphene. In figure S1, the parameters $t_t$ and $t_b$ correspond to the top and the back insulator thicknesses, respectively. The charge density $\rho_{\text{free}}(x,y)$ is zero inside both dielectrics so its only contribution corresponds to $\sigma_{\text{free}}(y)$ inside graphene. When solving the Poisson's equation, the electrostatic potential on the top gate and back gate is set to $V_{gs} - V_{gs0}$ and $V_{bs} - V_{bs0}$, respectively [2]. $V_{gs0}$ and $V_{bs0}$ are the flatband voltages that correspond to each gate. Homogeneous Neumann's conditions are applied to the other two boundaries to ensure charge neutrality [3]. At this stage we neglect the contribution of the supporting layer of hBN since the error we are making in evaluating the total gate capacitance remains below 3% for the values used in this work. This is due to the fact the hBN layer presents capacitance in series with a thick SiO$_2$ layer whose capacitance is much lower.

The drift-diffusion equation for the drain current $I_{ds}$ reads as follows:

$$I_{ds} = qW[n(y) + p(y)]\mu(y)\frac{dV}{dy}$$ (S2)

where $\mu(y)$ is the mobility of electrons and holes, and $V(y)$ is the quasi-Fermi potential in the graphene. The boundary conditions make $V(y)$ to be zero at $y = 0$ and $V_{ds}$ at $y = L$. Electron and holes share the same quasi-Fermi level due to a very short recombination time of carriers in graphene, of around 10 - 100 ns [4,5].

From the linear dispersion relation of graphene [6], and thus accounting for its quantum capacitance, the carrier concentrations are calculated by the following equations from both the electrostatic and quasi-Fermi potentials [1,7]:

$$n(y) = \frac{n_{\text{puddle}}}{2} + N_G\mathcal{F}_1\left[q\frac{\psi(0,y)-V(y)}{kT}\right]$$ (S3a)

$$p(y) = \frac{n_{\text{puddle}}}{2} + N_G\mathcal{F}_1\left[q\frac{V(y)-\psi(0,y)}{kT}\right]$$ (S3b)

We have added the contribution of graphene puddles $n_{\text{puddle}}$ to the carrier concentrations [8]. Here, $k$ is the Boltzmann constant, $T$ is the temperature (which is taken to be 300 K) and $N_G$ is the effective density of states of graphene, given by:

$$N_G = \frac{2}{\pi}\left(\frac{kT}{\hbar v_F}\right)^2$$ (S4)

being $\hbar$ the reduced Planck's constant and $v_F$ the Fermi velocity ($10^8$ cm s$^{-1}$). In equation (S3), $F_1(x)$ refers to the first order Fermi-Dirac integral:

$$\mathcal{F}_i(x) = \frac{1}{\Gamma(i+1)}\int_0^\infty \frac{u^i du}{1+e^{u-x}}$$ (S5)

The high field mobility model that we have used in this work includes saturation velocity effects in the following form:

$$\mu = \frac{\mu_{LF}}{\left\{1+\left|\frac{\mu_{LF}\partial\psi}{v_{sat}\partial y}\right|^\beta\right\}^{\frac{1}{\beta}}}$$ (S6)

where the parameter $\beta$ is roughly 2 for both electrons and holes, consistently with experiments [9]. By the same token, both the low-field carrier mobility $\mu_{LF}$ and saturation velocity $v_{sat}$ have been extracted by the Monte Carlo methodology discussed in section 2.2. These two properties of carriers in graphene depend on the charge concentration.

In summary, given the set of materials and geometry of the GFET, and after selecting a bias point ($V_{gs}$, $V_{bs}$ and $V_{ds}$), the simulator solves in a self-consistent way the drift-diffusion transport equation (S2) coupled with the 2D Poisson's equation (S1). The simulator then obtains the stationary values of $I_{ds}$, $n(y)$, $p(y)$, $\psi(x,y)$ and $V(y)$ as the outputs.

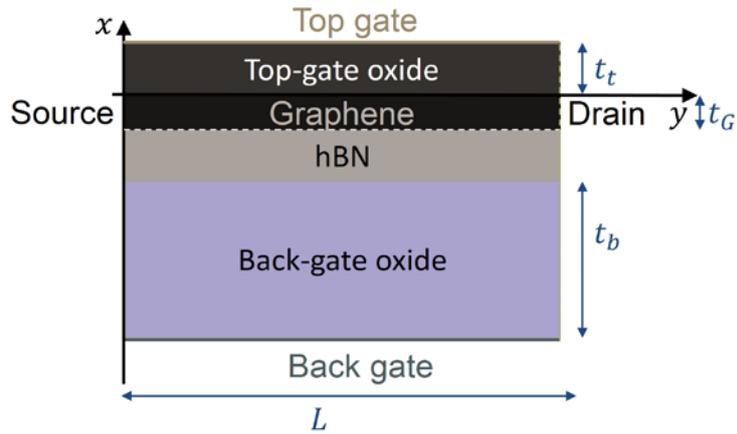

**Figure S1** Cross section of the GFET and the domain where the Poisson's equation is evaluated. This area corresponds to the dashed rectangle in figure 1 of the main text.

## S2. Small-signal parameters determination

This section explains how the small-signal matrix $Y(\omega)$ of the GFET is obtained from the stationary model explained in section S1. We consider here the two-port network represented in figure 3(a) of the main text and we assume that the back gate has a negligible influence over the graphene charge, which is reasonable if the back gate capacitance is much smaller than the top gate capacitance. The charge controlled by the gate, source and drain terminals ($Q_G$, $Q_S$ and $Q_D$, respectively) can be obtained from the self-consistent simulator after evaluation of the sheet charge distribution $\sigma_{\text{free}}(y)$. Upon application of a Ward-Dutton's linear charge partition scheme as the charge control model, the terminal charges read as [10]:

$$Q_D = W \int_0^L \frac{y}{L} \sigma_{\text{free}}(y) dy \tag{S7a}$$

$$Q_S = W \int_0^L \left(1 - \frac{y}{L}\right) \sigma_{\text{free}}(y) dy \tag{S7b}$$

$$Q_G = -W \int_0^L \sigma_{\text{free}}(y) dy \tag{S7c}$$

Notice that the total charge in the device is zero, so the model is charge-conserving. From the charge model described above, the intrinsic capacitances can be determined in the following way:

$$C_g = \left.\frac{\partial Q_G}{\partial V_{gs}}\right|_{V_{ds}} \tag{S8a}$$

$$C_{dg} = -\left.\frac{\partial Q_D}{\partial V_{gs}}\right|_{V_{ds}} \tag{S8b}$$

$$C_{sd} = -\left.\frac{\partial Q_S}{\partial V_{ds}}\right|_{V_{gs}} \tag{S8c}$$

$$C_{gd} = -\left.\frac{\partial Q_G}{\partial V_{ds}}\right|_{V_{gs}} \tag{S8d}$$

$$C_{gs} = C_g - C_{gd} \tag{S8e}$$

To complete the model, the transconductance $g_m$ and output conductance $g_d$ need to be evaluated.

$$g_m = \left.\frac{\partial I_{ds}}{\partial V_{gs}}\right|_{V_{ds}} \tag{S9a}$$

$$g_d = \left.\frac{\partial I_{ds}}{\partial V_{ds}}\right|_{V_{gs}} \tag{S9b}$$

As can be deduced from the diagram depicted in figure 3(b), the intrinsic admittance matrix then takes the form:

$$Y'(\omega) = \begin{bmatrix} j\omega C_g & -j\omega C_{gd} \\ g_m - j\omega C_{dg} & g_d + j\omega\left(C_{gd} + C_{sd}\right) \end{bmatrix} \tag{S10}$$

We must include the influence of the series resistances $R_G$, $R_S$ and $R_D$ to obtain the extrinsic admittance matrix $Y(\omega)$. According to figure 3(b), the relation between the extrinsic AC small-signal voltages $v_{gs}$ and $v_{ds}$ and the intrinsic ones, $v'_{gs}$ and $v'_{gs}$, can be written as:

$$v_{gs} = v'_{gs} + i_g(R_G + R_S) + R_S i_d \qquad (S11a)$$

$$v_{ds} = v'_{ds} + R_S i_g + i_d(R_D + R_S) \qquad (S11b)$$

The phasors $i_d$ and $i_g$ are, respectively, the output and the input small-signal currents of the two-port network as defined in figure 3(a). Then, we define the series resistance matrix $Z_c$ as:

$$Z_c = \begin{bmatrix} R_G + R_S & R_S \\ R_S & R_D + R_S \end{bmatrix} \qquad (S12)$$

Finally, from the equation S11 it can be deduced that the admittance matrix can be calculated by means of the following relation:

$$Y(\omega) = \{[Y'(\omega)]^{-1} + Z_c\}^{-1} \qquad (S13)$$

### S3. Monte Carlo model

The low-field mobility and saturation velocity are calculated by numerically solving the Boltzmann transport equation by means of the Monte Carlo method. We consider the energy-momentum dispersion relation to be conically shaped nearby the Dirac points, expressed as $\varepsilon(\mathbf{k}) = \hbar v_F |\mathbf{k}|$, where we take the Fermi velocity $v_F = 10^6$ m/s. This is a close approximation within the electric fields considered since a very limited fraction of the total particles in the simulation reach energies that imply significant deviations from more accurate band structure models.

The scattering rates are computed for each mechanism $m$ as a function of the initial carrier energy as

$$\lambda_m(\varepsilon_0) = \frac{\Omega}{4\pi^2} \int_{\mathbf{k}'} \frac{2\pi}{\hbar} |H_m(\mathbf{k_0}, \mathbf{k}')|^2 \Delta_m(\varepsilon, \varepsilon') \, \mathrm{d}^2\mathbf{k}' \tag{S14}$$

where $\mathbf{k_0}$ is the initial wavevector, that corresponds to the energy $\varepsilon_0$, $\mathbf{k}'$ is the final wavevector after the scattering takes place, $H_m(\mathbf{k_0}, \mathbf{k}')$ is the matrix element of the transition $\mathbf{k_0} \to \mathbf{k}'$ due to scattering with the mechanism $m$ and

$$\Delta_m(\varepsilon, \varepsilon') = \left(N_{\mathbf{q}} + \frac{1}{2} \pm \frac{1}{2}\right) \delta(\varepsilon' + \varepsilon' \mp \hbar\omega_{\mathbf{q}}). \tag{S15}$$

**Intrinsic phonon scattering**

Scattering with transversal and longitudinal acoustic (TA/LA) phonons with long wavelength (close to point Γ in the band structure) is treated as a quasi-elastic collision (it has been assumed that the phonon energy, $\hbar\omega_{\mathbf{q}}$ is negligible) and it is described by the first order deformation potential approximation. Its probability as a function of the electron energy can be obtained as [11,12]:

$$\lambda_{\text{TA/LA}-\Gamma}(\varepsilon_0) = \frac{\Omega}{4\pi^2} \int_{\mathbf{k}'} \frac{2\pi}{\hbar} \frac{\hbar D_1^2 |\mathbf{q}|}{2\Omega\rho_m v_{ph}} \left[1 - \left(\frac{|\mathbf{q}|}{2|\mathbf{k_0}|}\right)^2\right] \left(\frac{2k_B T_g}{\hbar v_{ph} |\mathbf{q}|}\right) \delta(\varepsilon - \varepsilon') \mathrm{d}^2\mathbf{k}' \tag{S16}$$

where $D_1$ is the deformation potential coupling constant, $\mathbf{q} = \mathbf{k} - \mathbf{k}'$ is the phonon wavevector corresponding to the difference of the initial and final states, $\rho_m$ is the two-dimensional graphene density, and $v_{ph}$ is the phonon velocity associated with the slope of the acoustic phonon dispersion at the Γ point.

As for the optical and acoustic inelastic modes, the zeroth-order deformation potential [12,13] is used assuming dispersionless ($\omega_{\mathbf{q}} \to \omega_{\mathbf{0}}$) phonons in the vicinities of the relevant critical points. The matrix element is given by:

$$\lambda_{ph}(\varepsilon_0) = \frac{\Omega}{4\pi^2} \int_{\mathbf{k}'} \frac{2\pi}{\hbar} \frac{\hbar D_0^2}{2\Omega\rho_m \omega_{\mathbf{0}}} \left(N_{\mathbf{q}} + \frac{1}{2} \pm \frac{1}{2}\right) \delta(\varepsilon_0 - \varepsilon' \mp \hbar\omega_{\mathbf{0}}) \mathrm{d}^2\mathbf{k}' \tag{S17}$$

In this expression, $D_0$ is the deformation potential, $N_{\mathbf{q}}$ accounts for the phonon occupation given by the Bose-Einstein statistics, and $\pm$ accounts for phonon emission/absorption.

In both models, the deformation potentials are set so that scattering rates calculated with the deformation potential models reproduce those obtained with *ab-initio* methods based on the density functional theory (DFT) [13]. The relevant parameters for these scatterings are given in table S1.

**Table S1** Inelastic phonon scattering.

|  | $v_{ph}$ ($10^4$ m/s) | $D_1$ (eV) | $D_0$ ($10^9$ eV/cm) | $\hbar\omega_0$ (meV) |
|---|---|---|---|---|
| TA-KA (Γ) | 2.0 [35] | 6.8 [35,36] | -- | -- |
| TA-LA (K,K') | -- | -- | 0.35 [13] | 124.0 [13] |
| TO-LO | -- | -- | 1.0 [13] | 164.6 [13] |

Some authors have pointed out that DFT within the local-density (LDA) and generalized-gradient (GGA) approximations may present some inaccuracies as compared to Green's functions based (GW) corrections related to the higher optical branch and Kohn anomalies at the symmetry points [14]. In our Monte Carlo model, electron-phonon scattering is treated in a common way in the framework of electronic transport, *i. e.*, by considering analytical expressions and adequate deformation potentials to reproduce the global effect of electron-phonon couplings. As indicated by Fischetti *et al.* [15] there is a lack of consensus in the literature regarding deformation potential values for graphene, due to multiple reasons related to the consideration of effective single modes, differences in the band structure models, etc. Moreover, as also pointed out in that paper, the deformation potentials are averaged over large regions of the Brillouin Zone, and necessarily correspond to effective deformation potentials that usually lump different modes into a single effective one. The values chosen here correctly reproduce the velocity-field curves of pristine graphene from [16].

**Surface optical phonons**

The interaction of free electrons in graphene with remote phonons from the substrate is treated by means of an additional scattering mechanism. Its scattering probability is given by [17,18]:

$$\lambda_{\text{SPP}}(\varepsilon_0) = \frac{\Omega}{4\pi^2} \int_{\mathbf{k}'} \frac{2\pi}{\hbar} 2\pi e \mathcal{F}^2 \frac{\exp(-2|\mathbf{q}|d)}{\Omega \bar{\varepsilon} \epsilon(|\mathbf{q}|,\omega)|\mathbf{q}|} \frac{1+\cos\theta_{\mathbf{k}\mathbf{k}'}}{2} \left(N_{\mathbf{q}} + \frac{1}{2} \pm \frac{1}{2}\right) \cdot$$
$$\cdot \delta(\varepsilon_0 - \varepsilon' \mp \hbar\omega_s) \, d^2\mathbf{k}'$$
(S18)

where $\mathcal{F}$ is the Fröhlich coupling constant, which is equal to:

$$\mathcal{F} = \left[\left(\frac{1}{\epsilon_\infty+1} - \frac{1}{\epsilon_0+1}\right)\frac{\hbar\omega_s}{2\pi\Omega}\right]^{1/2},$$
(S19)

$d$ is the Van der Waals distance between the graphene sheet and the surface of the substrate whose phonons provoke the scattering, $\mathcal{F}$ is the Fröhlich coupling constant, $\epsilon_\infty$, $\epsilon_0$ are the optical and DC substrate relative dielectric permittivities and $\hbar\omega_s$ is the energy of each surface polar phonon mode. The values of the parameters employed in the simulations are summarized in table S2.

**Table S2** Surface optical phonons.

| Substrate | $\epsilon_0$ | $\epsilon_\infty$ | $\hbar\omega_{s1}$ (meV) | $\hbar\omega_{s2}$ (meV) |
|---|---|---|---|---|
| hBN [37,38] | 5.09 | 4.10 | 101.70 | 195.70 |
| Al$_2$O$_3$ [23] | 12.53 | 3.20 | 55.01 | 94.29 |

The model accounts for screening via the static screening function $\epsilon(q, \omega \to 0) = 1 + V(q)\Pi(\mu, T_e, q)$, where $V(q)$ is the bare Coulomb potential, and $\Pi(\mu, T_e, q)$ is the electronic temperature-dependent polarizability, calculated within the random phase approximation [19]:

$$\Pi(\mu, T_e, q) = \frac{g_s g_v}{2\pi^2 \hbar v_F}\left\{\frac{\pi q}{8} + \frac{\mu + 2 k_B T_e \log\left[1+\exp\left(-\frac{\mu}{k_B T_e}\right)\right]}{\hbar v_F} - \int_0^{\frac{q}{2}}\left[\frac{1}{1+\exp\left(\frac{\varepsilon(k)+\mu}{k_B T_e}\right)} + \frac{1}{1+\exp\left(\frac{\varepsilon(k)-\mu}{k_B T_e}\right)}\right]\sqrt{1-\left(\frac{2k}{q}\right)^2} \, dk\right\}$$
(S20)

During the simulation, the electronic temperature $T_e$, and chemical potential $\mu$, are computed consistently with the system conditions by solving the following system of equations that relate these quantities with total carrier density and the average thermal energy:

$$n_e - n_p = \frac{2}{\pi}\left(\frac{k_B T_e}{\hbar v_F}\right)^2 \left[\mathfrak{F}_1\left(\frac{\mu}{k_B T_e}\right) + \mathfrak{F}_1\left(-\frac{\mu}{k_B T_e}\right)\right]$$
(S21a)

$$\langle\varepsilon_{th}\rangle = 2 \, k_B T_e \, \mathfrak{F}_2\left(\frac{\mu}{k_B T_e}\right)/\mathfrak{F}_1\left(\frac{\mu}{k_B T_e}\right)$$
(S21b)

The effect of the image charges that would be introduced in the interface of the metallic gate and the top dielectric may also modify the dielectric environment. In order to include this effect, we include a modification of the screened Coulomb scatterer via a Green's function as described in [20]. However, we found that for top-oxide thicknesses thicker than 15 nm this effect vanishes and, for the 26 nm-thick oxide in the present work its influence is negligible.

The static approximation to describe the polarizability has been frequently employed in the literature (see, for example [21–25]). However, it must be taken with caution. Some authors [26,27] indicate that the interaction between graphene plasmons and surface polar phonons results in the hybridization of these modes into the so-called interfacial-plasmon phonon (IPP) modes. According to ref. [27], as compared to statically screened SPP the main discrepancies with IPP modes appear in the very short and very long wavelength limits. In our case, very short wavelengths are extremely restricted due to the anisotropic nature of SPP interactions. On the other hand, small q transitions are limited by the SPP formulation. Consequently, our model may present some inaccuracies when dealing with the microscopic features of the interaction with remote phonons from the substrate and dynamic hybridization of the modes for very long wavelength transitions. However, since the main goal of the paper is to describe qualitatively the effect of impurities and defects on the performance of GFETs, we think that the approach chosen provides a reasonable approach to account for the substrate influence. Moreover, our model successfully reproduces experimental velocity-field curves for graphene on $SiO_2$ when impurities and defects are considered, and shows a good agreement with the results of more elaborate models for the treatment of remote-phonon interactions including hybridized modes [28].

**Carrier-carrier collisions**

The model accounts for short-range carrier-carrier scattering by means of the screened Coulomb potential associated to the transitions of the binary electron system $\mathbf{k_1} \rightarrow \mathbf{k_1'}$ and $\mathbf{k_2} \rightarrow \mathbf{k_2'}$ [29]:

$$V_{c-c}(\mathbf{k_1}, \mathbf{k_1'}, \mathbf{k_2}, \mathbf{k_2'}) = \frac{2\pi e^2}{\Omega \bar{\varepsilon} \varepsilon(|\mathbf{q}|, \omega)|\mathbf{q}|} \frac{1+\cos\theta_{\mathbf{k_1}, \mathbf{k_1'}}}{2} \frac{1+\cos\theta_{\mathbf{k_2}, \mathbf{k_2'}}}{2} \tag{S22}$$

where $\bar{\varepsilon}$ is the background dielectric constant [30], $\varepsilon(|\mathbf{q}|, \omega)$ is the dielectric function. Momentum-dependent scattering probability reads:

$$\lambda_{c-c}(\mathbf{k_1}) =$$
$$\left(\frac{\Omega}{4\pi^2}\right)^2 \int_{\mathbf{k_2}} \int_{\mathbf{k_1'}} \frac{2\pi}{\hbar} f(\mathbf{k_2}) \frac{1}{2} (|V_{c-c}(\mathbf{k_1}, \mathbf{k_1'}, \mathbf{k_2}, \mathbf{k_2'})|^2 +$$
$$|V_{c-c}(\mathbf{k_1}, \mathbf{k_2'}, \mathbf{k_2}, \mathbf{k_1'})|^2 +$$
$$|V_{c-c}(\mathbf{k_1}, \mathbf{k_1'}, \mathbf{k_2}, \mathbf{k_2'})V_{c-c}(\mathbf{k_1}, \mathbf{k_2'}, \mathbf{k_2}, \mathbf{k_1'})|^2)\, \delta(\varepsilon_1 + \varepsilon_2 - \varepsilon_1' -$$

<div style="text-align:right">(S23)</div>

$\varepsilon_2')\delta(\mathbf{k}_1 + \mathbf{k}_2 - \mathbf{k}_1' - \mathbf{k}_2')\, \mathrm{d}^2\mathbf{k}_1'\mathrm{d}^2\mathbf{k}_2$

### Charged impurities

Short-range Coulomb scattering with charged impurities is also included in the model. The scattering matrix is obtained from the 2D Fourier transform of the electrostatic potential. Considering that the impurities are distributed homogeneously on the material with a characteristic density $n_{imp}$, the scattering probability is [31]:

$$\lambda_{\mathrm{imp}}(\varepsilon_0) = \frac{\Omega}{4\pi^2} \int_{\mathbf{k}'} \frac{2\pi}{\hbar} n_{imp} \frac{2\pi e^2 \exp(-2|\mathbf{q}|d)}{\Omega \bar{\varepsilon} \varepsilon(|\mathbf{q}|,\omega)|\mathbf{q}|} \delta(\varepsilon - \varepsilon')d^2\mathbf{k}' \tag{S24}$$

Here, $n_{imp}$ is the impurity density, $d$ is the Van der Waals distance to the substrate, where the charged impurities are supposed to be on the substrate surface.

### Scattering with defects

Along with impurities, atomic scale defects play a crucial role in the electronic transport of fabricated samples. Both point defects and dislocations can be treated as a single scattering mechanism according to the approximation described in [32]. The treatment is similar to the case of impurity scattering, but the 2D transform of the electrostatic potential is replaced by a constant term. Finally, the defect scattering probability is:

$$\lambda_{\mathrm{def}}(\varepsilon_0) = \frac{\Omega}{4\pi^2} \int_{\mathbf{k}'} \frac{2\pi}{\hbar} N_{def} \left| \frac{V_0 L_{\mathrm{def}}^2}{\Omega} \right|^2 \frac{1+\cos\theta_{\mathbf{k},\mathbf{k}'}}{2}^2 \delta(\varepsilon - \varepsilon')d^2\mathbf{k}' =$$

$$\frac{\Omega}{4\pi^2} \int_{\mathbf{k}'} \frac{2\pi}{\hbar} \frac{\alpha^2}{\Omega} \frac{1+\cos\theta_{\mathbf{k},\mathbf{k}'}}{2}^2 \delta(\varepsilon - \varepsilon')d^2\mathbf{k}' \tag{S25}$$

where $N_{def}$ is the number and density of defects, $V_0$ is the energy associated to the average defect potential, $L_d$ is the effective potential range that relates to the size of the puddles formed in graphene, and $\alpha^2 = \sqrt{n_{def}}|V_0 L_{def}^2|$, being $n_{def}$ the density of defects.

## S4. Influence of the scattering types on all mobility scenarios

From the Monte Carlo data, it is possible to unravel the influence of each separate scattering type on the low field mobility and the saturation velocity. The results for the three scenarios are shown in figure S2, where the percentage of each scattering type is shown as a function of the carrier concentration.

In the lowest mobility case, scenario 3, the dominant mechanism is impurity scattering. As the carrier concentration grows, scattering with defects increases its weight, becoming the second interaction type together with interactions with remote surface polar phonons (SPP) from the top $Al_2O_3$. At high electric field, while impurity is still the dominant scattering type, interactions with the dielectrics and intrinsic optical phonons become also relevant at high $n_s$.

The second mobility scenario (intermediate mobility) shows also a dominant role of impurities and interactions with SPP of the top substrate: in particular, these latter ones become the most important scattering type at medium and large carrier concentrations. With regard to the saturation velocity conditions, the interactions with the dielectrics are the most relevant scattering type at medium and large carrier concentrations.

Finally, in the highest mobility case (absence of impurities or defects) the low and high field situations are mostly dominated by the interactions with remote SPP from $Al_2O_3$, although interactions with the underlying hBN are also relevant. A higher influence of intrinsic phonons is also observed as the carrier concentration increases.

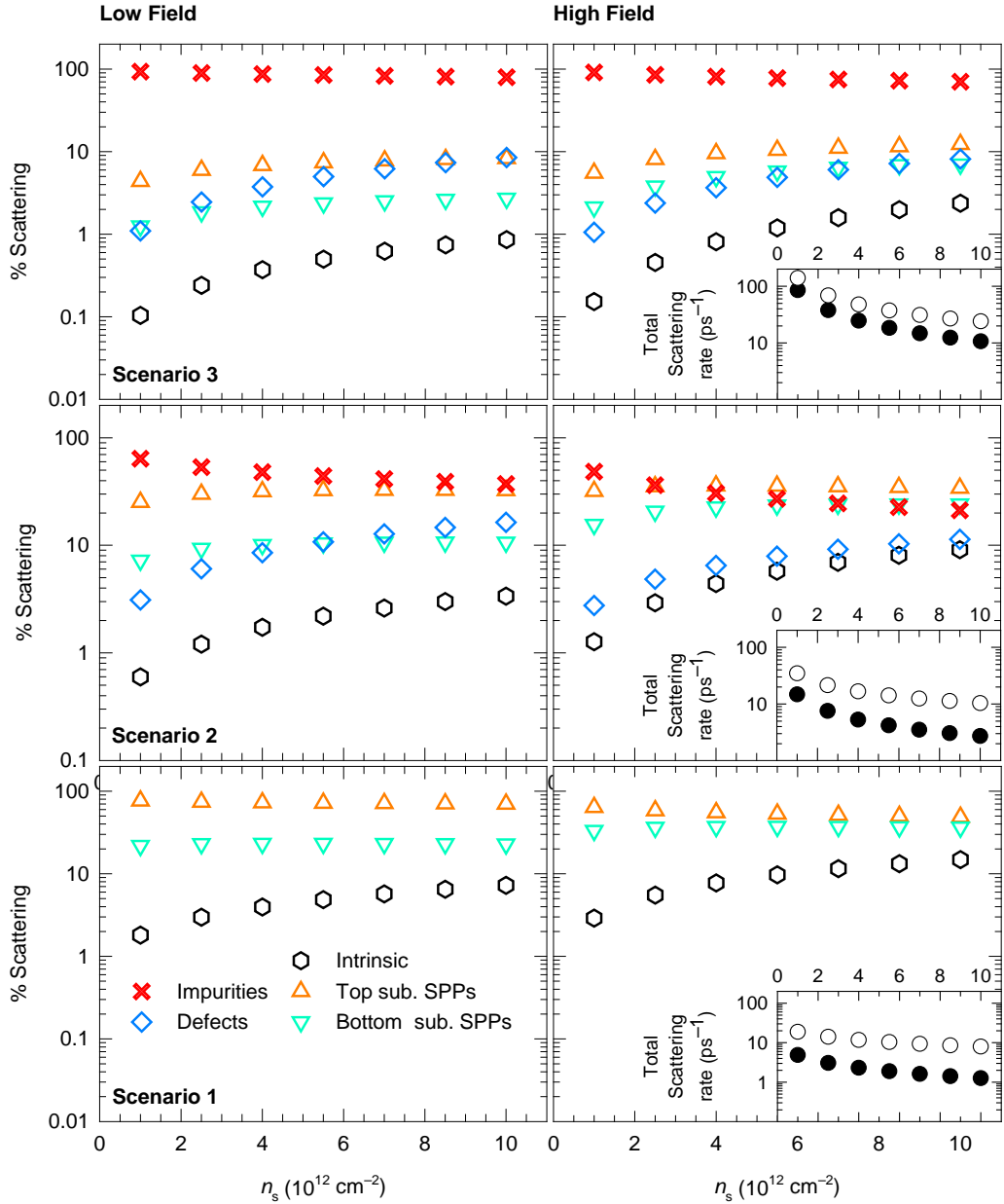

**Figure S2** Percentage of scatterings (left: low field; right: high field) in the three different mobility scenarios considered. Insets include the total scattering rate as a function of the carrier concentration at low and high field (black and white symbols, respectively).

## S5. Analytical curves for the mobility and saturation velocity derived from Monte Carlo simulations

From the Monte Carlo data for the mobility and saturation velocity, we have fitted the results to analytical curves to be included in the GFET simulator. For this purpose, in the case of the low field mobility we considered the following expression:

$$\mu_{LF} = \frac{\mu_0}{1+\left(\frac{n_s}{a_1}\right)^{b_1}} \tag{S26}$$

where $n_s$ is the carrier concentration and $\mu_0$, $a_1$, and $b_1$ are the fitting parameters. The fitting has been made in the $5\cdot10^{11}$-$10^{13}$ cm$^{-2}$ range.

Regarding the saturation velocity, it can be fitted to a potential law of the form:

$$v_{sat} = v_{sat,0}\left(1 + a_2\left(\frac{n_s}{n_0}\right)^{b_2}\right) \tag{S27}$$

where $v_{sat,0}$, $a_2$ and $b_2$ are the fitting parameters, and $n_0$ is a reference carrier concentration of $10^{12}$ cm$^{-2}$. The whole set of fitting parameters for the three scenarios are summarized in Table S3, and the results are shown in figure 2(a-b) in the main text.

Below a minimum carrier concentration of $5\cdot10^{11}$ cm$^{-2}$, both $\mu_{LF}$ and $v_{sat}$ have been considered to be constant, taking the value of both magnitudes at $5\cdot10^{11}$ cm$^{-2}$. The Monte Carlo extracted values are then used in equation (S6). The magnitudes of $\mu_{LF}$ and $v_{sat}$ have been simulated only for the electrons; the same values have been considered for holes, considering the symmetry in the conduction and valence bands in graphene close to the Dirac point.

**Table S3** Fitting parameters for electron mobility and saturation velocity to match the curves obtained from the Monte Carlo simulations.

|  | $n_{imp} = 8\cdot10^{12}$ cm$^{-2}$ $\alpha = 0.10$ eV nm (Scenario 3, mobility corresponding to the experimental values) | $n_{imp} = 9.5\cdot10^{11}$ cm$^{-2}$ $\alpha = 0.07$ eV nm (Scenario 2, improved quality realistic case) | $n_{imp} = 0$ cm$^{-2}$ $\alpha = 0$ eV nm (Scenario 1, defect and impurity-free graphene) |
|---|---|---|---|
| $\mu_0$ | $2.4\cdot10^3$ cm$^2$ V$^{-1}$ s$^{-1}$ | $1.4\cdot10^4$ cm$^2$ V$^{-1}$ s$^{-1}$ | $3.1\cdot10^5$ cm$^2$ V$^{-1}$ s$^{-1}$ |
| $a_1$ | $6.9\cdot10^{12}$ cm$^{-2}$ | $2.3\cdot10^{11}$ cm$^{-2}$ | $2.0\cdot10^8$ cm$^{-2}$ |
| $b_1$ | 0.33 | 0.42 | 0.35 |
| $v_{sat,0}$ | $2.5\cdot10^7$ cm s$^{-1}$ | $9.2\cdot10^7$ cm s$^{-1}$ | $1.9\cdot10^8$ cm s$^{-1}$ |
| $a_2$ | $-6.1\cdot10^{-2}$ | $-0.43$ | $-0.67$ |
| $b_2$ | 0.69 | 0.20 | $8.6\cdot10^{-2}$ |

## S6. Discussion on the validity of the drift-diffusion transport assumption

The mean free paths (MFP) obtained with the Monte Carlo simulator are presented in figure S3. The MFPs have been computed by accounting for the average value of the absolute distance travelled in the direction of transport divided by the average total number of scattering events in stationary conditions. The values have been calculated for an applied field equal to the average electric field in the channel ($V_{ds}/L$), for gate lengths between 18 μm and 50 nm, with a $V_{ds}$ value equal to 0.6 V, as considered in the manuscript.

In figure S3 we also present the ratio between the gate length and the MFP in each case, as a function of $L$, since this is the relevant feature in order to discuss the diffusive character of transport and the validity of the drift-diffusion model. The results are presented for two carrier concentrations, $10^{12}$ and $10^{13}$ cm$^{-2}$. As it can be observed, the MPF ranges from a few nanometers to more than 200 nm depending on the carrier concentration and the electric field; however, they are always significantly smaller than the featured gate length for each case, so the validity of the diffusive model is guaranteed.

Carrier-carrier interactions play a very relevant role for all the electric fields considered, while acoustic phonons are also important at low fields; optical phonons contribute to reduce the mean free path at larger fields (*i. e.*, smaller gate lengths). Impurity scattering and defects are also major contributors to a reduction of the MFP as compared to a "clean" case, specially at small carrier densities. Only in the event of having smaller devices than those considered in the manuscript, or if considering an extremely reduced $V_{ds}$ (in the mV range for the channel lengths considered here) the MFP could become larger or comparable to the channel length, thus entering into the quasi-ballistic regime.

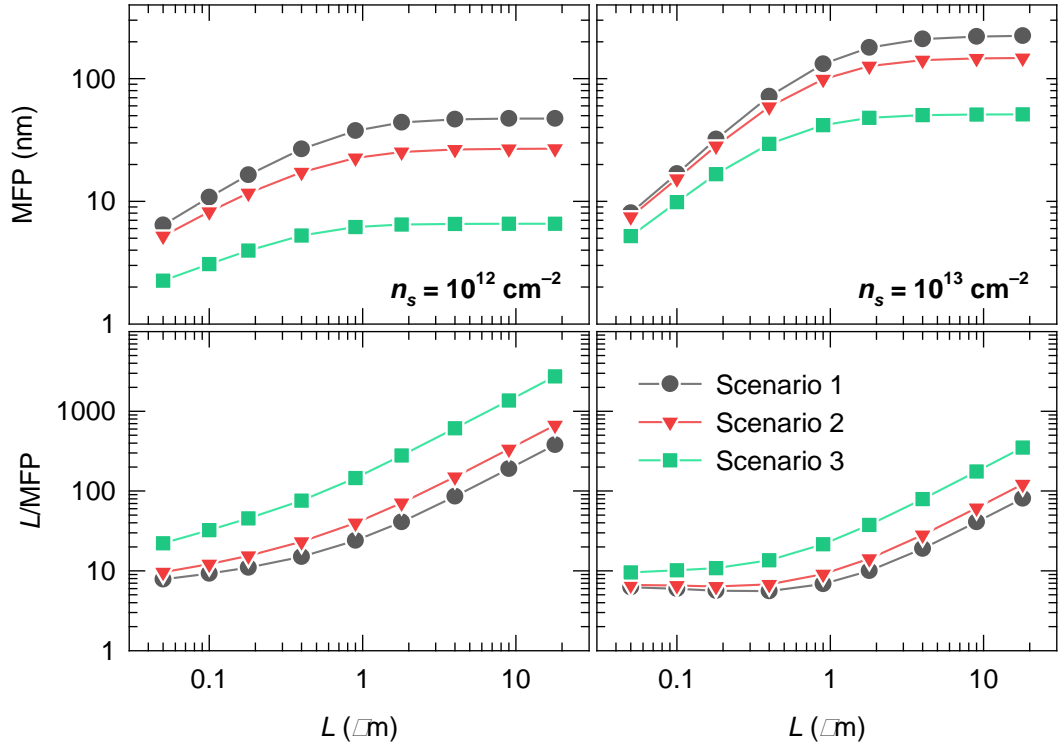

**Figure S3** Mean free paths (up) and channel length-to-MFP ratio (down) as a function of the gate length for two different carrier densities, $10^{12}$ cm$^{-2}$ (left) and $10^{13}$ cm$^{-2}$ (right).

### S7. Influence of interface traps

Although the simulated I-V curves reasonably agree with the experimental measurements, it is important to analyze the hypothetical effect that interface traps might have. Figure S4 compares the DC curves and the small-signal parameters of a GFET free from interface traps with a device assuming an interface trap capacitance $C_{it}$ of 10 fF µm$^{-2}$, which is a realistic value [33]. The traps in the graphene-dielectric interface decrease the number of carriers in the channel at a given bias, which decreases in turn the drain current and thus the transconductance and the output conductance (especially at biases that are far from the Dirac voltage). On the other hand, intrinsic capacitances barely remain unaffected.

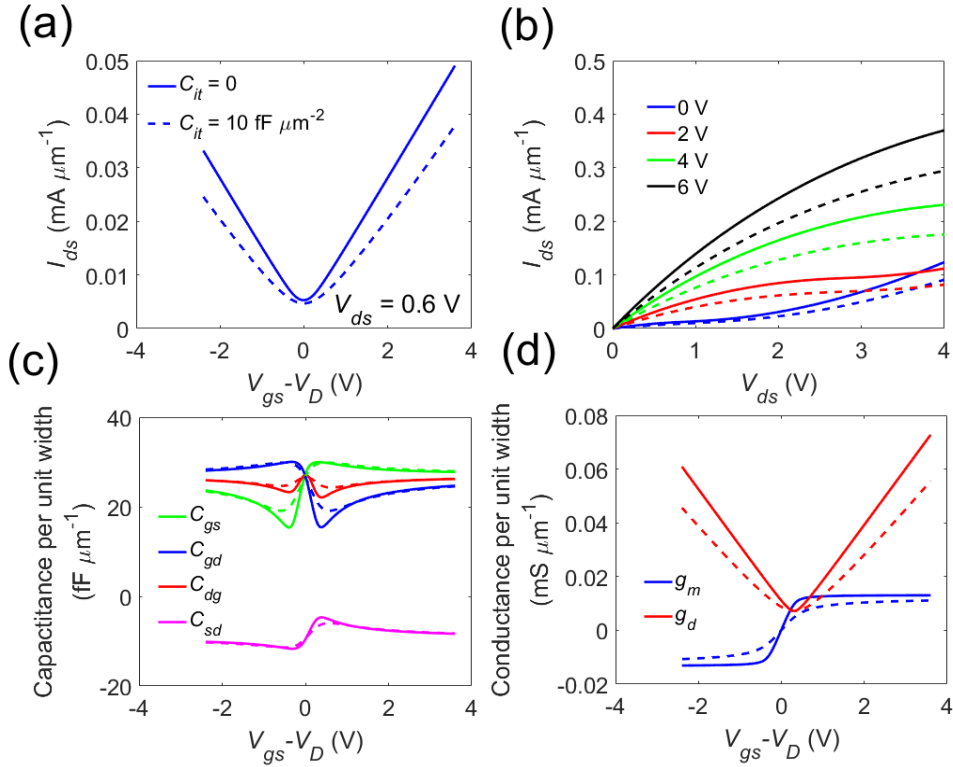

**Figure S4** Effect of the interface traps in the transfer characteristic at $V_{ds}$ = 0.6 V (a) and in the output characteristics at different $V_{gs}$ (b) for the nominal device (low mobility scenario and $L$ = 18 µm). Intrinsic capacitances (c) and transconductance and output conductance (d) at $V_{ds}$ = 0.6 V. Solid lines correspond to a trap-free interface while dashed lines to a $C_{it}$ of 10 fF µm$^{-2}$.

## S8. Influence of the puddles in RF performance

Figure S5 shows the effect of the carrier inhomogeneities (puddles) in the RF FoMs for different channel lengths. From figure S5(b), it can be observed that the density of puddles does not significantly affect $f_{T,x}$. However, $f_{max}$ slightly decreases as the puddle concentration increases, especially for short channels. This is caused by an increase of the output conductance with the presence of puddles, which is more noticeable in short-channel devices.

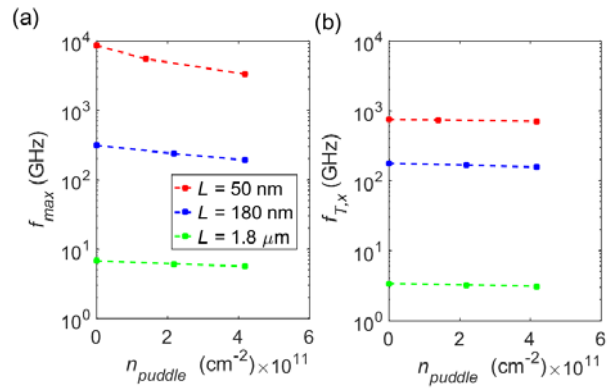

**Figure S5** $f_{max}$ (a) and $f_{T,x}$ (b) as a function of the puddle concentration

**S9. Choice of the bias point for RF performance investigation**

The GFET RF performances are, in general, dependent on the bias point. This can be seen in figure S6, which shows an exemplary plot of $f_{max}$ and $f_{T,x}$ at $V_{ds}$ = 0.6 V as a function of the gate voltage overdrive, $V_{gs}$ - $V_D$, where $V_D$ is the Dirac voltage. The graph exhibits two maxima and a minimum at $V_{gs}$ = $V_D$, both maxima happening when the channel is pinched-off whether at the source and drain sides, respectively, and the minimum when the channel is pinched-off just at the channel center. For our RF investigation, we have chosen the combination $V_{ds}$ = 0.6 V and $V_{gs}$ - $V_D$ = 2 V so the device is biased in the region where $f_{max}$ and $f_{T,x}$ are quite insensitive to $V_{gs}$. This makes the comparison of different devices easier.

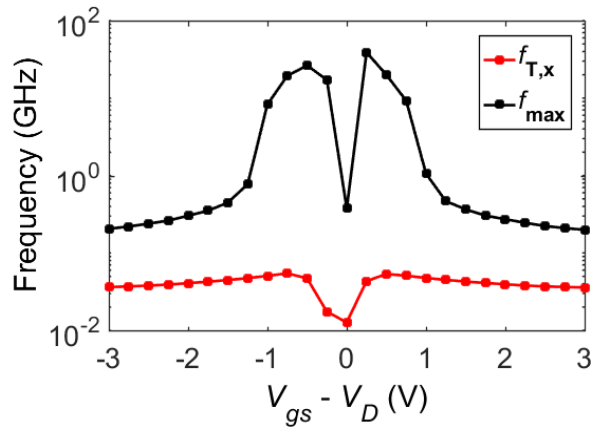

**Figure S6** Calculated extrinsic cutoff and maximum oscillation frequencies of the nominal device ($L$ = 18 μm, scenario 3 for mobility) as a function of the gate overdrive $V_{gs}$ - $V_D$. We have taken $V_{ds}$ = 0.6 V.

## S10. Scaling of small-signal parameters

To fully understand the scaling of $f_{max}$ and $f_{T,x}$, it is convenient to plot the behavior of the small-signal parameters of the GFET as a function of the channel length. Figure S7 shows how they scale for the high and low mobility scenarios. The transconductance scales as $1/L$ at long channels but it reaches a maximum value for $L < 1$ um, which is the reason for the observed saturation in the scaling of both $f_{max}$ and $f_{T,x}$. Regarding $g_d$, it can even reach negative values in short-channel devices (negative differential resistance). Its large drop causes $f_{max}$ to increase strongly, although its negative values may be the origin of the RF instability [34]. The intrinsic capacitances are approximately proportional to $L$ although the magnitudes of $C_{gd}$ and $C_{sd}$ strongly decrease at small $L$, especially in the case of high mobility.

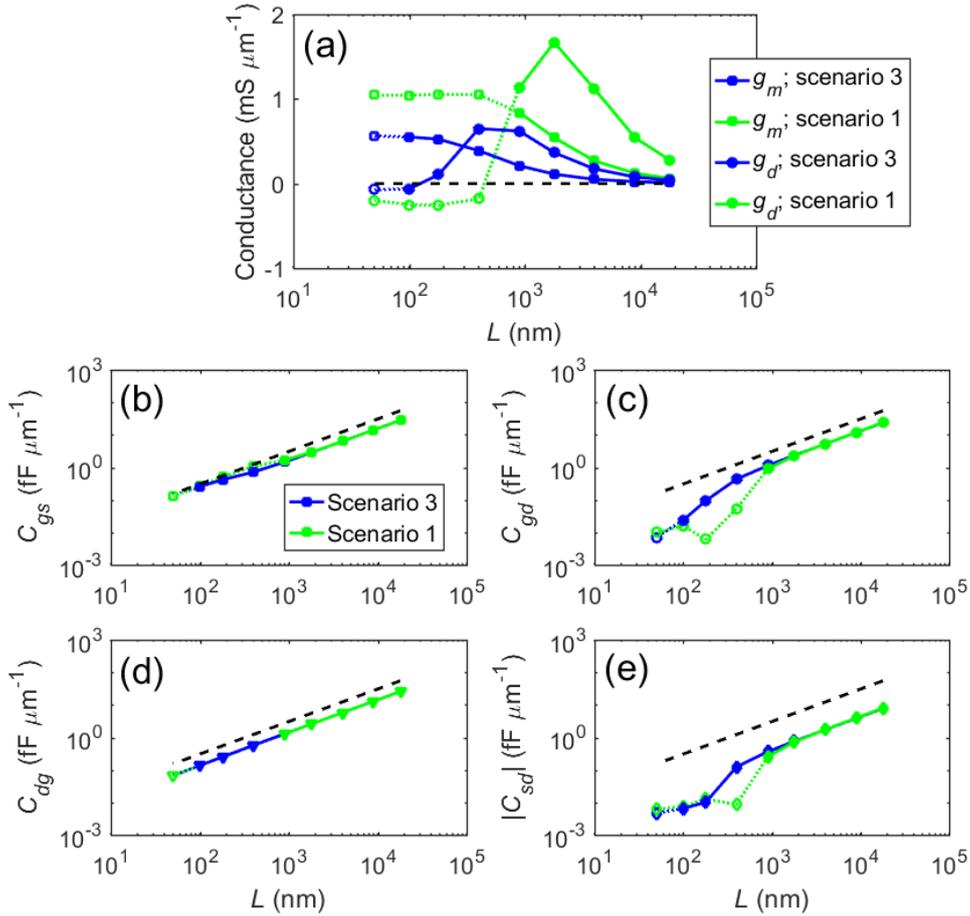

**Figure S7** Scaling of transconductance and output conductance (a), and intrinsic capacitances (b)-(e) per unit width. Dashed lines in capacitance graphs represent $C_t \cdot L$. The bias point in all cases is $V_{ds} = 0.6$ V and $V_{gs} - V_D = 2$ V.

**S11**. **Current-voltage characteristics**

Figure S8(a) compares the I-V characteristics of the 1.8 μm GFET with the experimental matched mobility, and a hypothetical GFET identical in geometry, but with the graphene channel just free from impurities and defects. The improvement in carrier mobility and saturation velocity increases $I_{ds}$ in a factor ~6. The transconductance $g_m$ of the GFET also grows while some current saturation is observed. These features indicate that the RF performance improves accordingly as we move towards the ideal scenario where graphene is free of defects and impurities (see figure 6(b) and (d) in the main text). On the other hand, figure S8(b) shows that scaling the channel length down to 180 nm produces an improvement in $g_m$ and a large increase of $I_{ds}$. Due to the increase of the electric field inside the channel, the drain current becomes limited by velocity saturation and the device saturates at a lower $V_{ds}$ (~0.5 V).

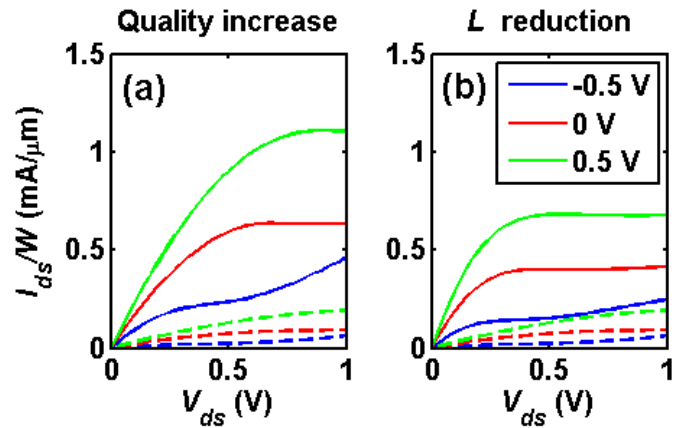

**Figure S8** (a) Simulated output characteristics of the device with $L$ = 1.8 μm (dashed lines) comparing the low mobility graphene (scenario 3, dashed lines) with a graphene free of defects and impurities (scenario 1, solid lines). (b) Simulated output characteristics of the 1.8 μm channel device (dashed lines) as compared to a GFET with a 180 nm channel (solid lines) for scenario 3. Curves are represented for $V_{gs}$ = -0.5, 0 and 0.5 V.